\pdfoutput=1
\documentclass{article}
\usepackage{spconf,amsmath,graphicx,hyperref,indentfirst}
\usepackage{diagbox}
\usepackage[dvipsnames]{xcolor}

\newlength{\bibitemsep}
\newlength{\bibparskip}
\let\oldthebibliography\thebibliography
\renewcommand\thebibliography[1]{%
  \oldthebibliography{#1}%
  \setlength{\parskip}{\bibitemsep}%
  \setlength{\itemsep}{\bibparskip}%
}

\hypersetup{pdfauthor={Hauret, Julien and Joubaud, Thomas and Zimpfer, Véronique and Bavu, Eric},
            pdftitle={EBEN},
            pdfsubject={Extreme bandwidth extension network applied to speech signals captured with noise-resilient body-conduction microphones},
            pdfkeywords={Speech enhancement} {PQMF-banks} {Bandwidth extension} {Frugal AI} {Signal Processing},}

\title{EBEN: Extreme bandwidth extension network applied to speech signals captured with noise-resilient body-conduction microphones}

\name{Julien Hauret$^{\star \dagger}$ \qquad Thomas Joubaud$^{\dagger}$ \qquad Véronique Zimpfer$^{\dagger}$ \qquad Éric Bavu$^{\star}$ }

\address{$^{\star}$ LMSSC, Conservatoire national des arts et métiers, Paris, France, HESAM Université  \\ $^{\dagger}$ Department of Acoustics and Soldier Protection, French-German Research Institute of Saint-Louis (ISL)}

\begin{document}

\maketitle

\begin{abstract}
In this paper, we present Extreme Bandwidth Extension Network (EBEN), a Generative Adversarial network (GAN) that enhances audio measured with body-conduction microphones. This type of capture equipment suppresses ambient noise at the expense of speech bandwidth, thereby requiring signal enhancement techniques to recover the wideband speech signal. EBEN leverages a multiband decomposition of the raw captured speech to decrease the data time-domain dimensions, and give better control over the full-band signal. This multiband representation is fed to a U-Net-like model, which adopts a combination of feature and adversarial losses to recover an enhanced audio signal. We also benefit from this original representation in the proposed discriminator architecture. Our approach can achieve state-of-the-art results with a lightweight generator and real-time compatible operation.
\end{abstract}


\vspace{-0.05cm}
\section{Introduction}
\label{sec:intro}

Speech capture for radio communications is traditionally performed using microphones located near the speaker's lips. However, this sound capture is sensitive to ambient noise, which reduces the intelligibility of communications in high noise levels such as in industry or on the battlefield. Under extreme conditions, even differential microphones are unable to get rid of high-level surrounding noise. Unconventional voice pickup systems such as bone conduction transducers, laryngophones or in-ear microphones integrated into hearing protections have great potential in many applications. These systems have higher impedance that is matched only by tissues and bones vibrations contrary to air molecules', making noise pollution almost imperceptible within captured speech \cite{bos2005speech,acker2005speech}.
Yet, further research is needed to optimize the effective bandwidth of the captured speech signal as mid and high frequencies are missing due to the intrinsic low-pass characteristic of the biological pathway.

Since the desirable system is a two-way communication device, this entails real-time execution constraints, \emph{i.e.} a short processing time ($\leq 20$~ms) to be indistinguishable to the human ear. Moreover, edge computations are required to guarantee low latency, which in turn necessitates lightweight architectures. These considerations also match frugal AI requirements. Finally, the developed model should be robust to gender, accent, and speech loudness.

To perform this audio super-resolution task, we opt for deep learning techniques since classical signal processing source-filter approaches \cite{iser2008bandwidth} have limited capabilities when the information has completely disappeared along a given frequency interval. Indeed, neural networks' ability to extract relevant features for the downstream task allows matching high and low frequency contents. Therefore, human processing should be minimal to let the network build its own representation, and could benefit from raw waveform inputs instead of spectrogram, mel-spectrogram\cite{davis1980comparison} or MFCC\cite{bogert1963quefrency}. This trend is endorsed by several works \cite{baevski2020wav2vec,germain2019speech,dai2017very} for various audio tasks. The use of raw audio can also be combined with multiband processing to speed up inference as in \textit{RAVE} \cite{caillon2021rave}. To pursue the objective of fast inference, a fully convolutional architecture has been preferred in \cite{oord2016wavenet,kuleshov2017audio}, eventually U-Net-like for audio-to-audio tasks \cite{stoller2018wave,bosca2021dilated}. In addition, a simple reconstruction loss may be insufficient for conditional generation, producing unrealistic samples. As shown in \cite{kumar2019melgan,kim2019bandwidth,kong2020hifi}, adversarial networks \cite{goodfellow2020generative} can significantly improve the naturalness of produced sound.

Based on the above observations, we developed EBEN, a new deep learning model, inspired by Seanet \cite{tagliasacchi2020seanet} to infer mid and high frequencies from speech containing only low frequencies. As in the original paper, we use a generator that maps the degraded speech signal to an enhanced version. The generator is optimized to produce samples which are close to the reference while maintaining a certain degree of naturalness at different time scales. A multiband decomposition using Pseudo-Quadrature Mirror Filters (PQMF) \cite{nguyen1994near} is applied to reduce the temporal dimension of input features and to focus the signal's discrimination solely on high frequency bands. We therefore share a common goal with \cite{li2021real}, but differ in methodology and addressed degradations.

We used recordings\footnote{\label{footnote:url}available to listen at \url{https://jhauret.github.io/eben} alongside source code, hyparameters and model performances} from a non conventional in-ear prototype in order to study the signal degradation. Focusing on capture-induced degradation of a specific device does not penalize the generality of our approach. This family of sensors consistently degrades speech in the same way, acting as a low-pass filter. Variations mostly occur in terms of cut-off frequency, attenuation, and lack of coherence at specific frequencies. Thus, to address any other system, it would be enough to have a matching dataset. The few minutes of recordings at our disposal being insufficient, we trained our model and several baselines \cite{kuleshov2017audio,kong2020hifi,tagliasacchi2020seanet,li2021real} with the French version of the Librispeech dataset \cite{pratap2020mls} with simulated degradations, hoping to later release a publicly available dataset of speech capture with noise-resilient body-conduction microphones.

\section{Methods}

\subsection{Degradation estimation}
\label{sec:degradation}

\begin{figure}[htb]
\begin{minipage}[b]{.48\linewidth}
  \centering
  \centerline{\includegraphics[width=4.0cm]{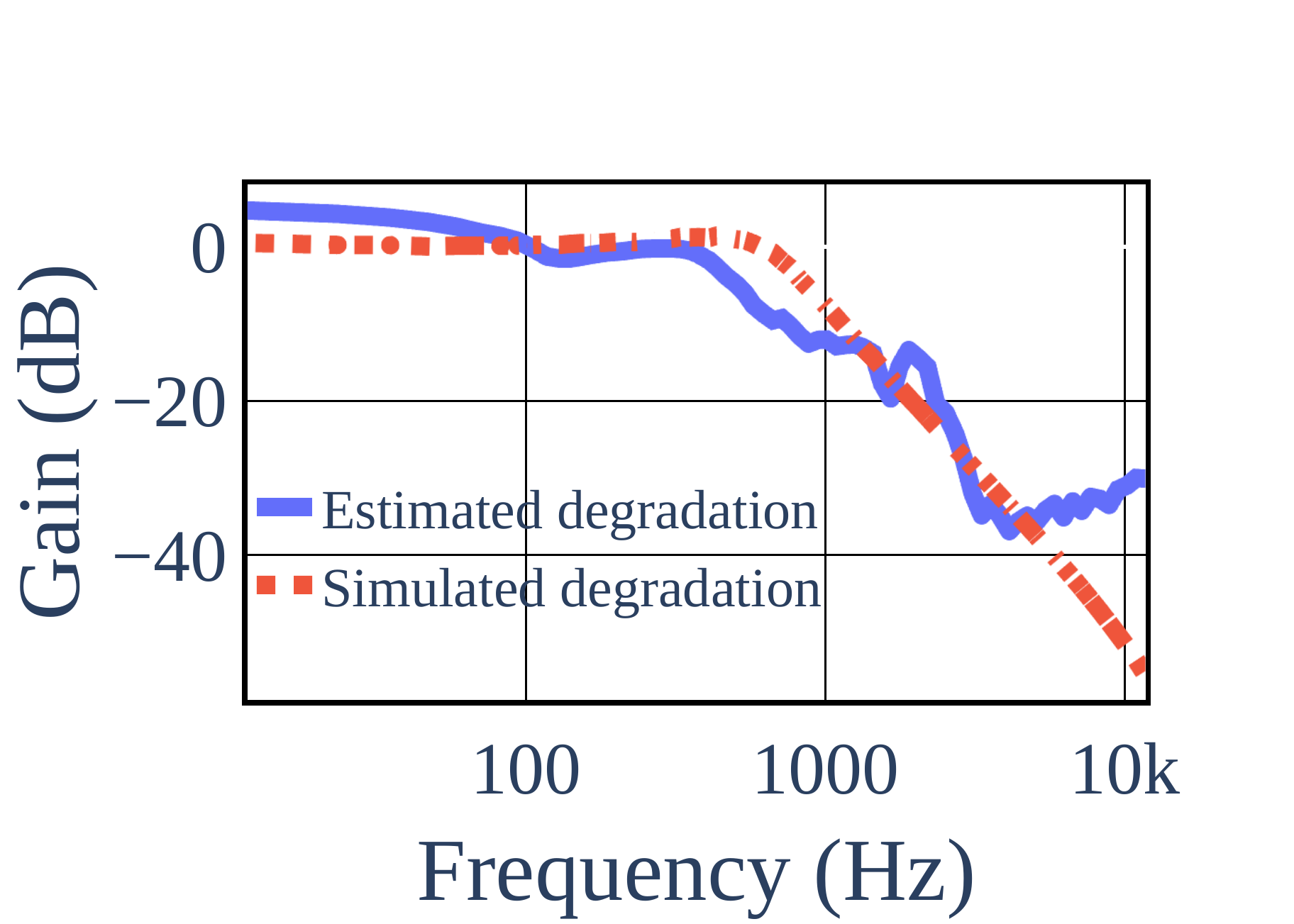}}
  \centerline{(a) Transfer function}\medskip
\end{minipage}
\hfill
\begin{minipage}[b]{0.48\linewidth}
  \centering
  \centerline{\includegraphics[width=4.0cm]{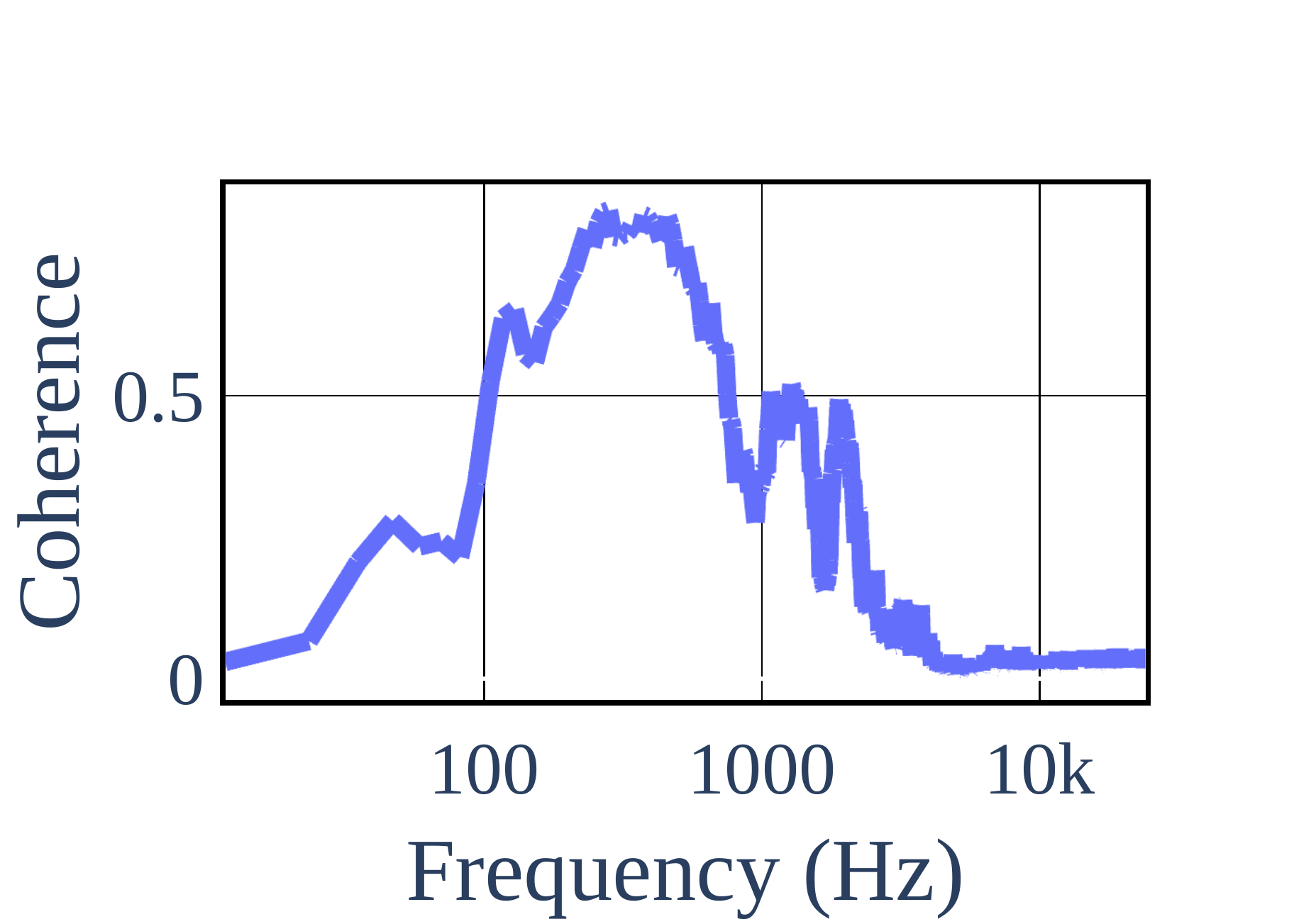}}
  \centerline{(b) Coherence function}\medskip
\end{minipage}
\caption{Degradation analysis of the in-ear transducer}
\label{fig:degradation}
\end{figure}

The selected equipment is an early prototype of in-ear transducer \cite{bionear} developed by ISL and Cotral Lab. This device takes advantage of the speaker's hearing protection by being placed
inside a custom-molded earplug, which increases communication capabilities in challenging and noisy environments. The captured signal mainly contains speech with no environmental noise. However, the acoustic path between the mouth and the transducers acts as a low pass filter: practically no relevant speech signal is picked up above a threshold frequency. This phenomenon can be further influenced by the occlusion effect due to the fitting of the individual protectors and some complex interactions with tissues. To evaluate this transfer function, the experimental protocol consists in capturing speech simultaneously with the in-ear transducer and a regular microphone placed in front of the speaker's mouth under noise-free conditions. Since speech signals are not stationary, several short-window estimates  were respectively used to produce an estimated degradation -- under the first approximation of a linear transfer function -- shown in Fig.~\ref{fig:degradation} (a) and the coherence function on Fig.~\ref{fig:degradation}(b).

The joint analysis of plots in Fig.~\ref{fig:degradation} shows that speech frequency content is only captured in a range below 2 kHz: the coherence function is close to zero above this frequency, and the gain of the transfer function is very low. This indicates that the in-ear microphone exhibits a very high attenuation at mid and high frequencies. Interestingly, at very low frequencies, the coherence function also indicates a lack of correlation between the two signals, since in-ear transducers also pick up physiological sounds produced by swallowing or blood flow. Finally, two anti-resonances are observed at 900 Hz and 1700 Hz, corresponding to vibration nodes of the occluded inner ear and propagation in bones and tissues.

\subsection{Pseudo Quadrature Mirror Filter formalism}
\label{sec:pqmf}

\begin{figure}[htb]
  \centering
  \centerline{\includegraphics[width=7.5cm]{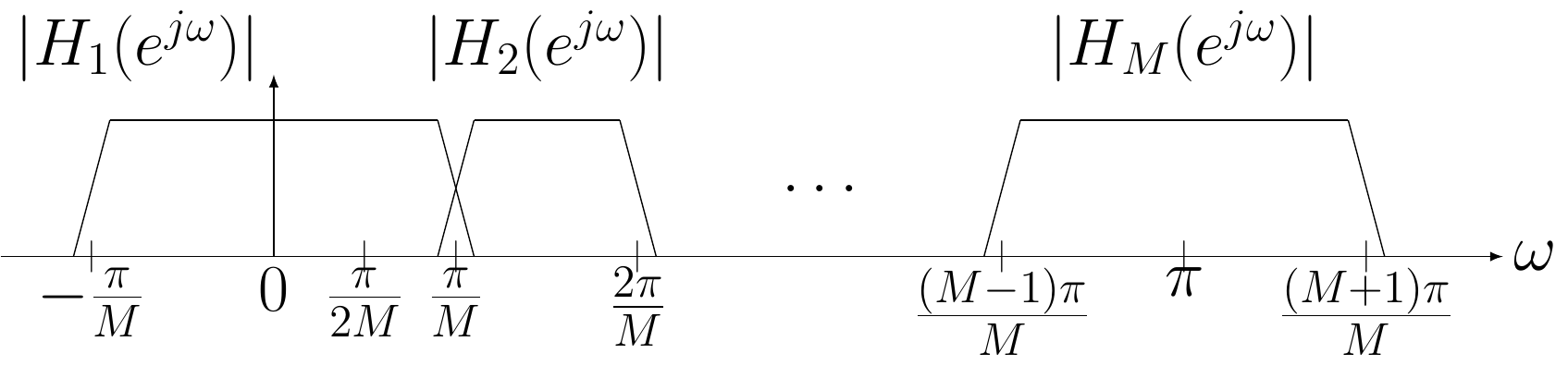}}
\caption{Frequency response of a PQMF ﬁlter bank}
\label{fig:bode}
\end{figure}

\begin{figure}[htb]
  \centering
  \centerline{\includegraphics[width=8.5cm]{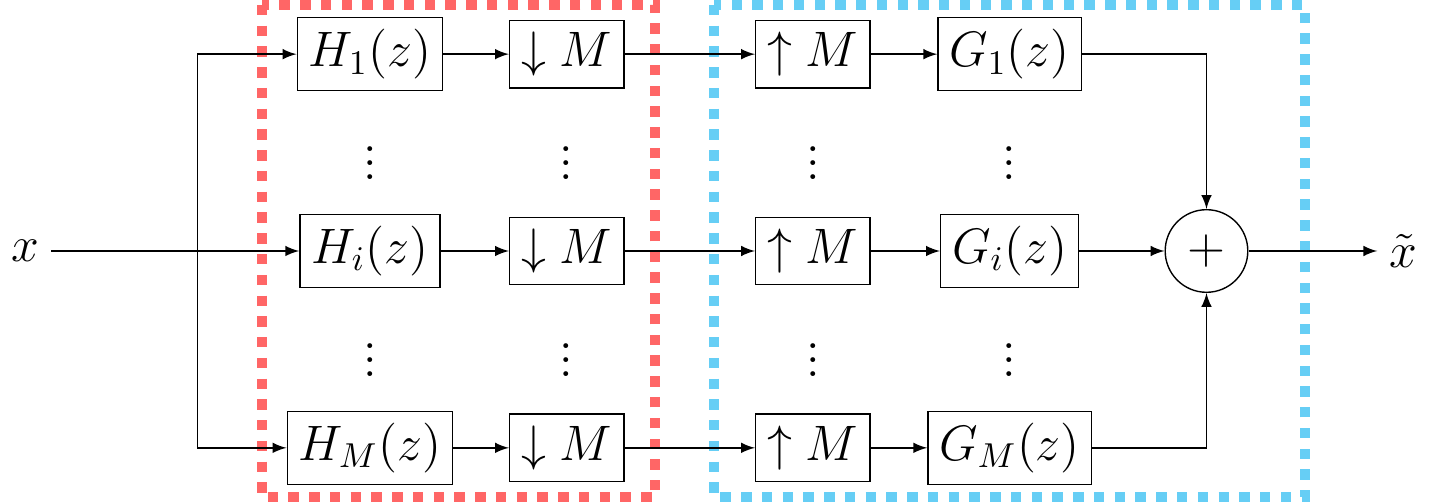}}
\caption{PQMF Analysis and Synthesis : block-diagram}
\label{fig:aands}
\end{figure}

The Quadrature Mirror Filter (QMF) banks, introduced in \cite{rothweiler1983polyphase}, are a set of analysis filters $\{H_i\}_{i \in [1,M]}$ used to decompose a signal into several non-overlapping channels of same bandwidth, and synthesis filters $\{G_i\}_{i \in [1,M]}$ used to recompose the signal afterward. Those filters are obtained from the same lowpass prototype filter.
A typical frequency response for a M-band PQMF bank is given in Fig.~\ref{fig:bode}. Fig.~\ref{fig:aands} shows the entire pipeline. The reconstruction is exact if $\{H_i\}_{i \in [1,M]}$ and $\{G_i\}_{i \in [1,M]}$ have an infinite support. In practice, a convolution kernel of $8M$ is enough to produce a near perfect pseudo reconstruction $\hat{x}$ with \textit{well-chosen}\footnote{with an optimization process to minimize the reconstruction error} filters \cite{nguyen1994near}.

 When used in a subband coding context, the PQMF analysis filter's outputs are decimated (downsampled) by a factor $M$. This can speed up inference because the different bands are modeled as conditionally independent. The PQMF analysis can also be seen as an downsampling operator, allowing to generate channels with considerably reduced redundancy, leading to a reduction in computational complexity. Along with EBEN source code, we provide a modern and efficient implementation of the PQMF analysis and synthesis with native Pytorch functions, using strided convolutions and strided transposed convolutions only.

\begin{figure*}[ht!]
  \centering
  \centerline{\includegraphics[height=3cm]{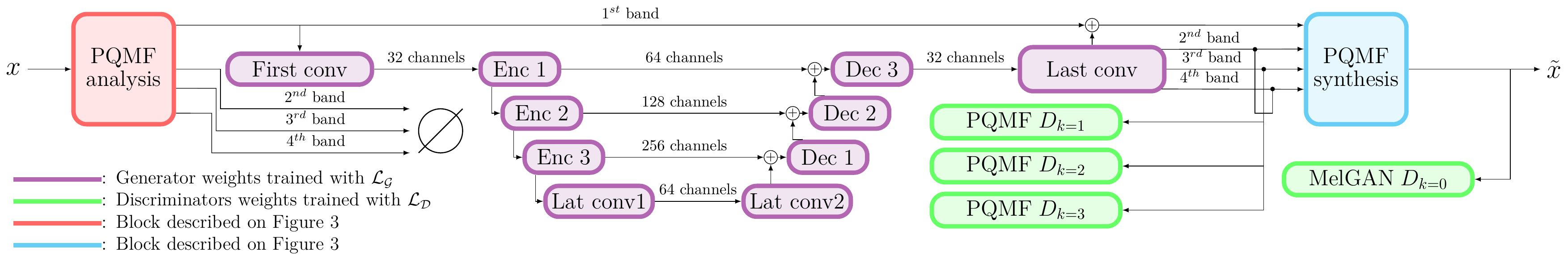}}
\caption{Overall architecture represented for $M=4$, one band with voice content fed to generator, others to PQMF discriminators}
\label{tikz:overall}
\end{figure*}

\subsection{Model architecture}
\label{sec:typestyle}

Unlike the Seanet generator, where the very first layer is a plain convolution used to process the audio at the original 16kHz sampling rate, we propose for EBEN to encapsulate a U-Net-like generator between a PQMF analysis layer and a PQMF synthesis layer. This allows to reduce the memory footprint of the model by reducing the first embeddings samplerate by a factor of M. It also makes possible to keep only subbands having voice content to feed the U-Net first convolution. Moreover, the number of encoder/decoder blocks is reduced to meet the constraints of real-time applications.

EBEN's discriminator also differs from the Seanet approach, as we directly exploit the PQMF subbands as multiple discriminator's inputs, without recombining nor upsampling the reconstructed subband signals. As shown in Fig~\ref{tikz:overall}, we adopt a multiscale ensemble discriminator approach, inspired by the work of Kumar \emph{et al.} in \cite{kumar2019melgan}, whose inputs are the upper bands of the PQMF decomposition. This allows the ensemble of discriminators to analyze the generated subband signals at different time scales, and helps the generator to be trained to generate high quality samples despite the fact that each discriminator is relatively simple. The subband discriminators $\{D_k\}_{k\in[1,2,3]}$ exhibit similar receptive fields to the original Seanet. As shown in Fig.~\ref{tikz:overall}, the full scale MelGAN discriminator was however kept from the Seanet approach, in order to ensure coherence between bands.

\subsection{Loss functions}
\label{sec:loss}

Discriminators are trained with a classical hinge loss. In $\mathcal{L_D}$, and $\mathcal{L_G}$, $G$ represents the generator and $D_{k,t}^{(l)}$ represents the layer $l$ of the discriminator (among $L_k$ layers) of scale $k$ (among $K$ scales) at time $t$. $F_{k,l}$ and $T_{k,l}$ are the number of features and temporal length for given indices. $x$ is the in-ear signal while $y$ is the reference.

\begin{equation*}
\begin{split}
 \mathcal{L_D}= E_y\left[ \frac{1}{K} \sum_{k \in [0,3]} \frac{1}{T_{k,L_k}} \sum_t max(0,1-D_{k,t}(y))\right] + \\ E_x\left[ \frac{1}{K} \sum_{k \in [0,3]} \frac{1}{T_{k,L_k}} \sum_t max(0,1+D_{k,t}(G(x)))\right]
 \label{eq:ld}
 \end{split}
\end{equation*}

For the generator, we adopted a loss function composed of a generative part $\mathcal{L_G}^{adv}$ and a reconstructive part $\mathcal{L_G}^{rec}$, balanced by $\lambda$ : $\mathcal{L_G}= \mathcal{L_G}^{adv} + \lambda \mathcal{L_G}^{rec}$.

\vspace{-0.15cm}

\begin{equation*}
\mathcal{L}_\mathcal{G}^{adv}= E_x\left[ \frac{1}{K} \displaystyle \sum_{k \in [0,3]} \frac{1}{T_{k,L_k}} \sum_t max(0,1-D_{k,t}(G(x)))\right]
\label{eq:lg_adv}
\end{equation*}

\vspace{-0.25cm}

\begin{equation*}
\mathcal{L}_\mathcal{G}^{rec}= E_x\left[ \frac{1}{K} \displaystyle \sum_{\substack{k \in [0,3] \\ l \in [1,L_k [ }} \frac{1}{T_{k,l}F_{k,l}} \displaystyle \sum_t \| D_{k,t}^{(l)}(y)-D_{k,t}^{(l)}(G(x))\| _{L_1}  \right]
\label{eq:lg_recons}
\end{equation*}

This combination allows to generate audio samples as close as possible to the reference signal thanks to $\mathcal{L}_\mathcal{G}^{rec}$, while remaining creative at high frequencies when no information is available in the degraded signal (especially for fricatives) thanks to $\mathcal{L}_\mathcal{G}^{adv}$. Note that $\mathcal{L}_\mathcal{G}^{rec}$ does not operate directly in the time domain but in discriminators domain to focus on the signal semantic (feature matching).

\section{Experiments}
\label{sec:experiments}

\begin{table*}[ht!]
    \centering
    \begin{tabular}{|l||l|l|l||l|l||l|l|}
     \hline
      \diagbox[width=8.5em, height=0.6cm]{Speech}{Metrics}  &  PESQ  &  SI-SDR  &  STOI & MUSHRA-U & MUSHRA-Q & Gen params & Dis params  \\
      \hline
      Simulated In-ear  &  \textbf{2.42 (0.34)}  &  8.4 (3.7)  &  0.83 (0.05)  &  51 (29)  &  24 (18)  &  $\emptyset$&  $\emptyset$ \\
      Audio U-net \cite{kuleshov2017audio}   &  2.24 (0.49)  &  \textbf{11.9 (3.7)}  & 0.87 (0.04)  &  60 (26) &  33 (18) & 71.0 M &  $\emptyset$\\
      Hifi-GAN v3\cite{kong2020hifi}   &  1.32 (0.16)  & -25.1 (11.4)  & 0.78 (0.04) &  40 (23)  &  36 (18) &  1.5 M &  70.7 M\\
      Seanet  \cite{tagliasacchi2020seanet} &  1.92 (0.48)  &  11.1 (3.0)  &  \textbf{0.89 (0.04)}  &  \textbf{73 (13)}  &  \textbf{78 (12)} &  8.3 M &   56.6  M \\
      Strm-Seanet  \cite{li2021real} &  2.01 (0.46) &  11.2 (3.6)  &  \textbf{0.89 (0.04)}  &  66 (20)  &  61 (14) &  \textbf{0.7 M} &   56.6  M \\
      EBEN (ours)   &  2.08 (0.45)  & 10.9 (3.3)  & \textbf{0.89 (0.04)} &  \textbf{73 (14)}  &  \textbf{76 (14)} &  1.9 M &  \textbf{27.8 M}\\
      \hline
    \end{tabular}
     \caption{Comparing models with: PESQ/SI-SDR/STOI on test set | MUSHRA-U (88 participants) and MUSHRA-Q (82 participants) scores | number of parameters. Format is median (IQR). Significantly best values (acceptance=0.05) are in \textbf{bold}.}
	\label{tab:comparison}
\end{table*}

\begin{figure*}[ht!]
  \centering
  \centerline{\includegraphics[width=18.5cm]{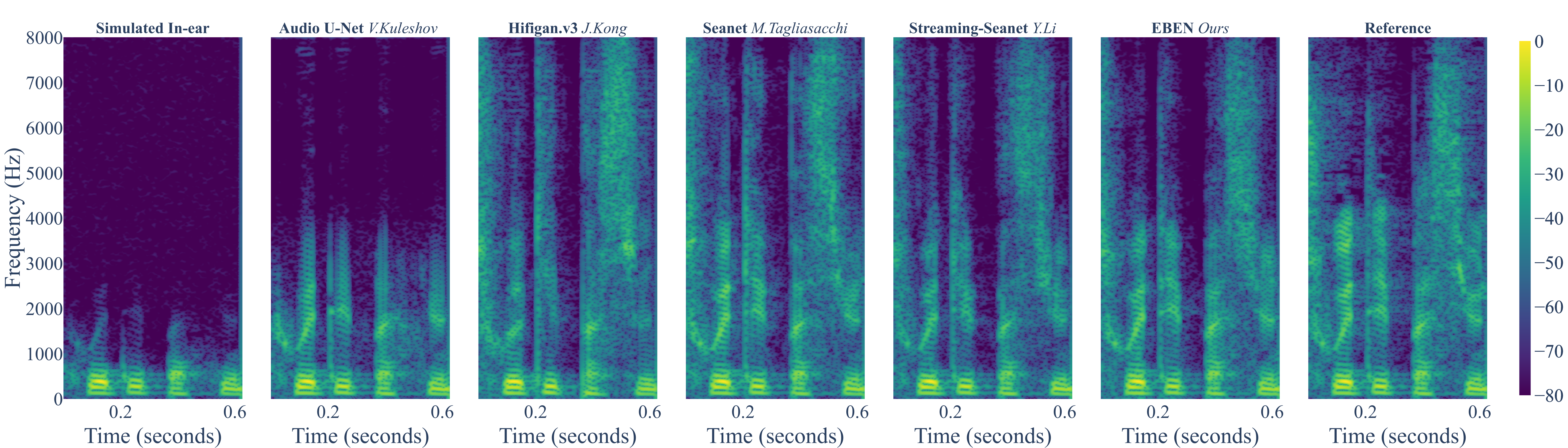}}
\caption{Spectrograms of various bandwidth extension models sandwiched by the simulated in-ear and the reference signals.}
\label{fig:spec}
\end{figure*}

A simulated degradation (described in Sec.~\ref{sec:simulation}) is performed on the French Librispeech \cite{pratap2020mls} dataset. Different models \cite{kuleshov2017audio,kong2020hifi,tagliasacchi2020seanet,li2021real} and the EBEN approach have been trained on this imitated in-ear dataset during 2 days on a single RTX 2080 Ti GPU. Each model is evaluated objectively (Sec.~\ref{sec:objective}) and subjectively (Sec.~\ref{sec:subjective}). No parameter tuning nor early stopping was performed.

\subsection{Simulation of the dataset}
\label{sec:simulation}
The simulated degradation is based on an autoregressive moving-average model, using a 2\textsuperscript{nd} order low-pass filter with a cutoff frequency of 600~Hz and unitary Q-factor that is applied using a \emph{filtfilt} procedure to ensure zero phase shift. The superposition of the simulated and estimated degradation filters are represented on Fig.~\ref{fig:degradation}~(a). A Gaussian white noise with a power 23 dB below the selected signal is also added.
This simplistic degradation still ensures a wide application field for developed algorithms and the ability to focus on the bandwidth extension issue. Future works will introduce variable filters with anti-resonance frequencies and realistic physiological noise.
\subsection{Objective evaluation}
\label{sec:objective}

To evaluate the model performances, Tab.~\ref{tab:comparison} highlights several objective metrics: Perceptual Evaluation of Speech Quality (PESQ) \cite{rix2001perceptual}, Scale-Invariant Signal-to-Distortion Ratio (SI-SDR)\cite{le2019sdr} and Short-Time Objective Intelligibility (STOI)\cite{taal2010short} on test set. Speech enhancement being a one-to-many problem, these results should be analyzed cautiously. Indeed, a plausible signal with perfect intelligibility but still different from reference would be misjudged by the metrics. Note that these metrics are intrusive, since they require a groundtruth audio. Generally speaking, speech quality assessment is lacking non subjective and non comparative evaluation metrics. Works like \textit{Noresqa} \cite{manocha2021noresqa} attempted to introduce such non intrusive metrics, but were inefficient for our specific degradation. We also report on Tab.~\ref{tab:comparison} the number of parameters for each model, showing that EBEN is among the lightest. Time latency and computational load are presented on the project website \ref{footnote:url}.

\vspace{-0.1cm}

\subsection{Subjective evaluation}
\label{sec:subjective}
To visually assess and compare the results of approaches, Fig.~\ref{fig:spec} shows some spectrograms obtained from the testing set. It can be observed that a purely reconstructive approach \cite{kuleshov2017audio} is not sufficient to produce high frequencies. Indeed, when low frequency information is insufficient, the model predicts the mean of speech signals which is zero. Among generative approaches, our method reconstructs a fair amount of formants and minimizes artifacts. We also conducted a subjective evaluation of the different methods using the MUltiple Stimuli with Hidden Reference and Anchor \cite{series2014method} (MUSHRA) methodology on the GoListen platform \cite{barry2021go}. Obtained results are also given on Tab.~\ref{tab:comparison}. MUSHRA-U is about ease of understanding while the MUSHRA-Q is about audio quality. Our approach is ranked first ex-aequo with Seanet for both aspects.

\vspace{-0.1cm}

\section{Discussion and Future Work}
\label{sec:conclusion}
EBEN is a realtime-compatible network architecture to address the problem of bandwidth extension of speech signals captured with body-conduction microphones. The proposed multiband approach offers several benefits, including reduced parameters and the ability to select the bands to operate on with a GAN approach. Developing a lightweight causal architecture able to capture long-range dependencies to disambiguate some phonetical content could help to further improve speech reconstructions for severe degradations. \\

\textbf{Acknowledgements}: This work has been partially funded by the French National Research Agency under the ANR Grant No. ANR-20-THIA-0002. This work was also granted access to the HPC/AI resources of [CINES / IDRIS / TGCC] under the allocation 2022-AD011013469 made by GENCI.

\bibliographystyle{IEEEbib}
{\small \bibliography{eben.bib}}

\end{document}